%% file: mimetas.tex
\documentclass[12pt,letterpaper]{article}

\usepackage{amsmath} 
\usepackage{amssymb} 
\usepackage{mathabx} 
\usepackage{siunitx} 

\usepackage[hypcap=false]{caption} 
\usepackage{graphicx} 
\usepackage{float} 
\usepackage{subfig}
\usepackage{tikz}
\usetikzlibrary{positioning}

\usepackage{booktabs} 
\usepackage{multirow} 

\usepackage{natbib} 

\usepackage[title]{appendix} 

\usepackage[inline]{enumitem} 
\usepackage{hyperref} 
\usepackage{parskip}
\usepackage{csquotes} 
\usepackage{lscape} 

\usepackage{authblk} 
\newcommand{\blind}{0}

\usepackage{orcidlink} 
\usepackage[version=3]{mhchem} 

\begin{document}
\def\spacingset#1{\renewcommand{\baselinestretch}%
{#1}\small\normalsize} \spacingset{1}

\if0\blind
{
\title{Design and analysis of a microplate assay in the presence of multiple restrictions on the randomization}
\author[1]{Alexandre Bohyn \orcidlink{0000-0001-9776-7467}}
\author[1]{Eric D. Schoen \orcidlink{0000-0002-4431-6117}}
\author[2]{Chee Ping Ng \orcidlink{0000-0002-4508-2770}}
\author[2]{Kristina Bishard}
\author[2]{Manon Haarmans}
\author[2]{Sebastian J. Trietsch}
\author[1,3]{Peter Goos \orcidlink{0000-0002-3854-6506}}
\affil[1]{Department of Biosystems, Faculty of Bioscience Engineering, KU Leuven, Leuven, Belgium}
\affil[2]{Mimetas BV, Leiden, The Netherlands}
\affil[3]{Department of Engineering Management, Faculty of Business and Economics,
University of Antwerp, Antwerp, Belgium}
\maketitle
}
\fi

\if1\blind
{
  \bigskip
  \bigskip
  \bigskip
  \begin{center}
    {\LARGE\bf Design and analysis of a microplate assay in the presence of multiple restrictions on the randomization}
\end{center}
  \medskip
} \fi

\bigskip
\begin{abstract}
\noindent
Experiments using multi-step protocols often involve several restrictions on the randomization.
For a specific application to {\em in vitro} testing on microplates, a design was required with both a split-plot and a strip-plot structure.
On top of two-level treatment factors and the factors that define the randomization restrictions, a multi-level fixed blocking factor not involving further restrictions on the randomization had to be added.
We develop a step-by-step approach to construct a design for the microplate experiment and analyse a response.
To consolidate the approach, we study various alternative scenarios for the experiment.

\end{abstract}

\noindent%
\vfill

\newpage
\spacingset{1.45} 

\section{Introduction}\label{sec:intro}
Empirical research and development projects frequently involve two-level screening designs to make out which factors in a design are active and which are not \citep[see the textbooks of, e.g.,][]{box1978statistics,mee2009comprehensive,wu2009experiments,montgomery2019design}. 
One of the strong points of these designs is their run size economy.
Indeed, it is feasible to include as many as $N-1$ factors in $N$ runs, where $N$ is a multiple of four.
A further strong point of two-level screening designs is their potential to include restrictions in the randomization and still allow for a simple analysis of the data.
There is a substantial body of literature on restrictions involving nested random factors.
These restrictions result in blocked designs \citep[see the aforementioned textbooks, as well as][]{sartono2015blocking}, row-column designs \citep{vo-thanh2020rowcolumn}, split-plot designs \citep{huang1998minimumaberration,bingham1999minimumaberration,bingham2001design,bingham2004designing,sartono2015constructing}, and split-split-plot designs \citep{schoen1999designing,jones2009doptimal}.
Other papers address strip-plot designs that involve crossed random factors \citep[see][]{miller1997stripplot,mee1998splitlot,vivacqua2004stripblock}.

The construction of designs in the presence of a combination of block restrictions, split-plot restrictions and strip-plot restrictions was addressed by \citet{bingham2008factorial} and \citet{cheng2011multistratum}.
However, it is not clear how to use their approaches when the design requires a multilevel fixed blocking factor, on top of the two-level treatment factors and the factors that define the randomization restrictions.
One such a case is the microplate experiment at the core of the present paper.
The experiment involved cells grown on plates with eight columns of wells.
There were eight two-level treatment factors.
Two weeks were reserved for the work, and one of the treatment factors was varied over the weeks.
In each week two plates could be processed; the second treatment factor was varied over the plates within a week.
Then, eight tubes with extra-cellular matter had to be prepared in each week, while each tube was used on each of the two plates for that week.
There were four factors that governed the contents of the tubes.
In addition, there were two treatment factors applied to the columns of a plate.
Finally, we needed to ensure that all the treatment factors were orthogonal to an eight-level factor indicating the column position on a plate. 

The purpose of this paper is to provide a step-by-step approach to design a fractional factorial experiment with multiple randomization restrictions and an extra multilevel factor.
In Section \ref{sec:problem}, we introduce the microplate experiment.
In Section \ref{sec:design}, we describe the protocol for the experimental conduct.
We then discuss the specific restrictions on the randomization of the experiment and outline a step by step approach to construct a suitable design for the experiment.
In Section \ref{sec:analysis}, we analyse the results of the experiment first by identifying active effects, and, second, by building a model that captures the full complexity of the error structure.
In Section \ref{sec:alternatives}, we revisit the step-by-step approach to consider four alternative scenarios of the experiment.
Finally, Section \ref{sec:conclusions}, offers some concluding remarks. 

\section{The microplate experiment}\label{sec:problem}

Organ-on-a-chip systems contain miniaturized tissues grown inside microfluidic chips.
They have rapidly gained popularity over the last decade because they mirror the morphology of \textit{in vitro} cells, which makes them more physiologically relevant than two-dimensional cell cultures \citep{sung2009control,vanduinen2015microfluidic,leung2022guide}.
However, until recently, the scalability of these systems was limited, because most setups use a single chip.
Furthermore, most systems use a permeable membrane for culturing cells, which can be an additional source of variation on top of the biological variation.
Recently, Mimetas BV, a company based in Leiden, The Netherlands, introduced the OrganoPlate$^{\circledR}$, a microfluidic multi-chip microplate where the cells are grown directly against an extra-cellular matrix (ECM), without the use of an artificial membrane. 

\begin{figure}[t]
\centering
\includegraphics[width=\textwidth]{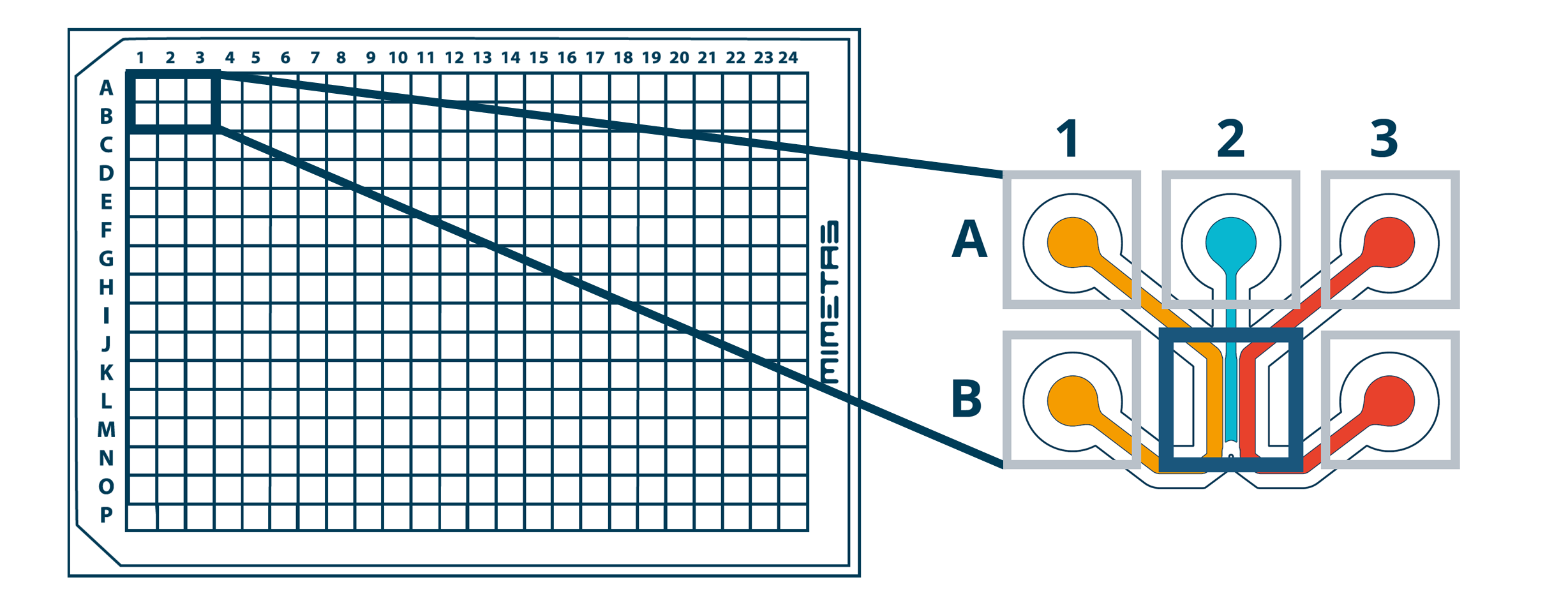}
\caption{Schematic representation of an OrganoPlate$^{\circledR}$ containing 64 chips. The inlay represents a single chip made up of six wells: (A1) left perfusion inlet, (B1) left perfusion outlet, (A2) gel inlet, (B2) observation window, (A3) right perfusion inlet and (B3) right perfusion outlet. Each chip contains three channels: a central extra-cellular matrix channel (blue), and two perfusion channels (red and orange).}
\label{fig:plate}
\end{figure}

The OrganoPlate$^{\circledR}$ consists of a microplate with 64 microfluidic chips, each consisting of six interconnected wells; see Figure \ref{fig:plate}.
The 64 chips are arranged in eight rows and eight columns.
The inlay shows that a single chip comprises three channels, the central ECM channel and two perfusion channels, joining in the center of the chip.
To prepare a plate for an experiment, the ECM is first prepared independently as a gel and then loaded on the chip through the gel inlet, A2.
Once the gel is set, experimenters may choose to study cells on both sides of the ECM or only on one side.
To this end, a medium containing cells is prepared and introduced in one or both of the perfusion channels of the chip through the perfusion inlets, A1 and A3 respectively.
Once the cells have fully grown, the plate is ready for testing.
All the measurements are made in the observation window of a chip, B2.
Each plate thus allows for 64 observational units.

The ECM is a complex network of macro-molecules created using a mix of different polymers.
The main polymer used in the ECM is type-I collagen because it allows the creation of unique geometries.
These geometries mimic three-dimensional (3D) tissues \textit{in vivo} and provide support to the surrounding cells \citep{theocharis2016extracellular}.
However, the physical properties of collagen are strongly influenced by its polymerization conditions, such as pH and temperature \citep{roeder2002tensile}.
In order to make this 3D system a reliable culture platform, it is essential that results are uniform within a plate, as well as reproducible between plates and between laboratories.
For this reason, Mimetas wants to detect the factors that affect the quality of their extra-cellular matrix.
In particular, the goal of the experimental study is to detect which factors minimize variation in the quality of the ECM.
A secondary goal specifically for the OrganoPlate$^{\circledR}$ was to establish factor settings that maximize that quality.

%

\section{Design construction}\label{sec:design}
In this section, we describe the protocol for preparing and loading the ECM gel and list the factors considered in this protocol.
Then, we detail the randomization restrictions in the experiment and explain how we built a suitable experimental design compatible with all these restrictions.

\subsection{Treatment factors}
We identified eight factors that could affect the quality of the ECM gel; they are shown in Table \ref{tab:factor_levels}.
The ECM gel is prepared by mixing several ingredients in an Eppendorf tube.
Before the actual mixing, the tube is cooled in an ice bath where it can stay for 1 or 30 minutes (factor $\mathbf{h}$).
The collagen solution that is added to the tube is taken from a large batch, which is aliquoted, i.e., divided into smaller batches.
The aliquot used can either come from the start or from the end of the original batch.
The origin of the aliquot is denoted by a two-level categorical factor with levels \enquote{Early} and \enquote{Late} (factor $\mathbf{a}$).
From each aliquot, either \SI{100}{\micro\litre} or \SI{300}{\micro\litre} of collagen can be added to the mix (factor $\mathbf{b}$).
Next, the mix is homogenized with a pipette, and this can be repeated either 20 or 50 times (factor $\mathbf{c}$).
The pH is known to influence the structure of collagen fibers and it can be adjusted by changing the \ce{NaHCO3} concentration in the basic solution.
The final pH of the gel is either 7.1 or 8.3 (factor $\mathbf{d}$).
When the preparation is finished, the mix needs to rest for 10 or 60 minutes (factor $\mathbf{g}$) before being pipetted into the plate.
Finally, once the gel has been loaded into the plate and has had time to polymerize, \SI{50}{\micro\litre} of Hanks' Balanced Salt Solution (HBSS), with either low or high levels of \ce{Ca++} and \ce{Mg++} (factor $\mathbf{e}$), is added in the gel inlet to prevent the gel from drying.
Before seeding the cells, the HBSS can either be removed or not.
This is studied using another two-level factor with levels \enquote{No} and \enquote{Yes} (factor $\mathbf{f}$).

\input{tables/factor_levels}

\subsection{Restrictions on the randomization}
To perform the experiment, four plates were available.
There are 64 chips per plate, resulting in a total of 256 individual chips.
For logistic reasons, the factors $\mathbf{g}$ and $\mathbf{h}$ could only be varied between the plates.
Since the ECM gel is prepared in an Eppendorf tube, factors $\mathbf{a,b,c}$ and $\mathbf{d}$ were all applied to an individual tube.
However, the ECM gel is loaded on the plate using a micro-pipette that fills an entire column of a plate at a time, so that these factors can only be varied over an entire column.
The same is true for factors $\mathbf{e}$ and $\mathbf{f}$, but since these factors are applied in the post-loading stage, they are not applied to individual tubes but to the columns of the plates.
Therefore, only eight different combinations of the factors $\mathbf{a}$ to $\mathbf{f}$ can be applied to a single plate, so that the experiment involves 32 different treatments combinations or experimental runs.

The manual processing of a single plate with many treatment combinations is time consuming.
Therefore, only two plates can be processed in a single week so that the experiment involves two weeks and the plates are nested within the weeks.
Since the ECM gel that is loaded into all chips within a column of a plate is prepared in an Eppendorf tube, a natural setup would have been to use 32 different tubes, one for each combination of column and plate.
However, only eight tubes could be prepared per week, so that each tube had to be used twice in a given week.
Our solution was to cross the tubes with the plates and nest them within the weeks.
We discuss the option of using the tubes twice on the same plate in Section \ref{sec:alt_scen_tube_twice}.
The diagram in Figure \ref{fig:structure_diagram} summarizes the experimental layout.

\begin{figure}[t]
    \centering
    \input{figures/structure_diagram}
    \caption{Schematic representation of the distribution of the tubes over the weeks and the plates. A black bullet in row $i$ and column $j$ means that the $j$th tube was used for the $i$th plate. Plates 1 and 2 are processed in week 1, plates 3 and 4 are processed in week 2.}
    \label{fig:structure_diagram}
\end{figure}

It was logistically convenient to vary factors $\mathbf{g}$ and $\mathbf{h}$ between entire plates, with one factor varied between the weeks and another varied between the two plates within a week.
Therefore, the four combinations of the levels of $\mathbf{g}$ and $\mathbf{h}$ define the four plates, and the main effect of $\mathbf{g}$, the main effect of $\mathbf{h}$, and the interaction effect $\mathbf{gh}$ are confounded with the random variation between plates or weeks.
This is unavoidable given the randomization restrictions.
Of the six remaining factors, four ($\mathbf{a}$ to $\mathbf{d}$) are varied between tubes, while the other two, $\mathbf{e}$ and $\mathbf{f}$, are varied between the columns of a plate.

In the design of experiments jargon, the fact that plates are nested within weeks and that one factor is applied to weeks and another is applied to  individual plates creates a split-plot design.
Here, the weeks are the whole plots, and within each whole plot, both the tubes and the plates are subplots.
On top of this, having the plates crossed with the tubes results in a strip-plot design in each of the two weeks.
Therefore, the design is a split-strip-plot design.

A common source of unwanted variability in microplate assays is the \enquote{edge effect}, due to which the measurements obtained from chips on the edge of a plate may differ from those obtained from chips in the center of the plate.
An edge effect can be caused by heterogeneous experimental conditions over the plate, such as a thermal gradient or a difference in the evaporation rate.
To minimize the impact of edge effects on the estimates of the factors' effects, we want to ensure that, over the four plates, the levels of the factors $\mathbf{a}$, $\mathbf{b}$, $\mathbf{c}$, $\mathbf{d}$, $\mathbf{e}$ and $\mathbf{f}$ are balanced across the eight column positions.
The experiment should therefore be blocked in eight blocks of size four.

\subsection{Treatment design}
\label{sec:trmt_design}
Our challenge is to create a 32-run design involving the eight two-level factors in Table \ref{tab:factor_levels}, and one eight-level blocking factor corresponding to the column positions within the plates, obeying all logistical restrictions.
To create the design, we proceeded in four steps:
\begin{enumerate*}[label=(\arabic*)]
    \item creating the design for the six factors $\mathbf{a}$ to $\mathbf{f}$ varied over the columns of a plate,
    \item arranging the design in eight blocks corresponding to the eight column positions of a plate,
    \item adding the two remaining factors $\mathbf{g}$ and $\mathbf{h}$ varied over the plates and the weeks, respectively, and
    \item ensuring that in each week, eight different tubes are used.
\end{enumerate*}

In step 1, we select the resolution-VI $2^{6-1}$ design for the six factors $\mathbf{a}$ to $\mathbf{f}$ with $\mathbf{f=abcde}$ as the generator to define the treatments for the factors varied between the columns of a plate.

In step 2, we arrange the 32 treatments of the $2^{6-1}$ design into eight blocks, such that the number of two-factor interactions that are confounded with the blocks is minimized.
The eight-level blocking factor can be defined by taking three independent factorial effects of the $2^{6-1}$ design. This confounds the three effects, the products of each pair of these effects, and the product of all three effects with the blocks, making up a total of seven confounded effects.
We call these seven effects \enquote{pseudo-factors}.
Choosing $\mathbf{ab}$, $\mathbf{ce}$ and $\mathbf{acf}$ (which is aliased with $\mathbf{bde}$) as the independent effects minimizes the number of two-factor interactions confounded with the blocks \citep{xu2006minimum}.
Table \ref{tab:aliasing-eight-level-factor} shows the aliasing of each of the seven resulting pseudo-factors with two- and three-factor interactions among the factors $\mathbf{a} - \mathbf{f}$.
The pseudo-factors $\mathbf{p_1}$, $\mathbf{p_2}$ and $\mathbf{p_3}$ correspond to the three independent factorial effects, $\mathbf{p_4}=\mathbf{p_1p_2}$, $\mathbf{p_5}=\mathbf{p_1p_3}$, $\mathbf{p_6}=\mathbf{p_2p_3}$ and $\mathbf{p_7}=\mathbf{p_1p_2p_3}$.
The column positions, determined by the grouping scheme presented in Table \ref{tab:grouping-scheme-eight-level}, are now orthogonally blocked for the main effects of factors $\mathbf{a}-\mathbf{f}$.

\input{tables/aliasing-eight-level-factor}

\input{tables/grouping-scheme}

In step 3, we use the OA package of \cite{eendebak2019oapackage} to generate all possible ways of adding an orthogonal four-level factor to the design with one eight-level blocking factor and six two-level factors, with the four levels of the new factor corresponding to the four plates and thus to the four level combinations of factors $\mathbf{g}$ and $\mathbf{h}$.
We obtained three different designs where the effects of the six two-level factors are either orthogonal to or completely confounded with the new four-level factor.
Table \ref{tab:aliasing-four-level-factor} shows the assignments that minimize the aliasing of the main effects of factors $\mathbf{g}$ and $\mathbf{h}$ with the two- or three-factor interactions among the other factors.
Option 1 is the best of the three because none of the main effects of $\mathbf{g}$ or $\mathbf{h}$ are aliased with a two-factor interaction.

\input{tables/aliasing-four-level-factor}

In the final step of the design creation, we need to assign each of the 16 possible level combinations of factors $\mathbf{a} -  \mathbf{d}$ to the 16 tubes available and ensure that eight of the 16 tubes are allocated to one week and the rest to the other week.
The design in Step 1 includes all 16 level-combinations of the factors $\mathbf{a} -  \mathbf{d}$, so that the experiment requires 16 tubes.
To define the weeks we can either use the factor $\mathbf{g=ace+bdf}$ or the factor $\mathbf{h=abc+def}$.
If we use factor $\mathbf{g}$, we do not invoke further restrictions on the tube factors $\mathbf{a}-\mathbf{d}$, so that each of the 16 tubes is used once each week.
However, if we use factor $\mathbf{h}$, the eight tubes for which $\mathbf{abc}=-1$ are used twice within the first week, and the eight remaining tubes are used twice within the second week.
Thus, using the factor $\mathbf{h}$ to identify the weeks and $\mathbf{g}$ to identify the plates within a week results in the required eight tubes per week.
This is shown in Table \ref{tab:tube-factor-aliasing}.
Finally, the full experimental design is presented in Table \ref{tab:full_design_table}.

Based on the $2^{6-1}$ design for factors $\mathbf{a}$ to $\mathbf{f}$,  and the aliasing of the pseudo-factors $\mathbf{p_i}$ $(i=1,\ldots,7)$ and the factors $\mathbf{g}$ and $\mathbf{h}$ with the factors of the $2^{6-1}$ design, we can compute the aliasing between all the main effects, the two-factor interactions and the pseudo-factors corresponding to the column positions.
The full aliasing pattern is presented in Table \ref{tab:aliasing}.
Furthermore, since we know the allocation of the treatments to the tubes, the plates, and the weeks, we can group the factorial effects according to their treatment unit and thus according to the random error they are subjected to.
The resulting groups are called \enquote{strata} and there are four of them in this experiment: $Week$, $Plate$, $Tube$, and $Unit$.
Both $Plate$ and $Tube$ are nested within $Week$, while $Plate$ is crossed with $Tube$.
Finally, $Unit$ corresponds to the smallest treatment unit, the column within a plate.

\input{tables/tube-factor-aliasing}

\renewcommand*{\arraystretch}{0.8}
\input{tables/full_design_table}
\renewcommand*{\arraystretch}{1}

\input{tables/total_aliasing}

\section{Analysis and Interpretation}\label{sec:analysis}
In this section, we describe the response variable used in the analysis.
Next, we present our analysis strategy.
Since we have eight factors, a model with an intercept, all main effects and all interactions would contain 37 terms.
However, only 32 runs are available.
For this reason, we have to run our analysis in two steps.
We first identify the active effects in each error stratum.
Next, we use the active effects to build the final model and interpret the results.

\subsection{Response variable}
\label{sec:response_variable}
An important characteristic of the ECM in the central channel of a chip is the fibrosity of the collagen network in the matrix.
To assess the fibrosity of the ECM in this experiment, a small sample of the ECM was collected in the central channel of each chip, and its fibrosity was measured using texture analysis \citep{mostaco-guidolin2013collagen}, that results in a dimensionless response.
A high measured value implies a low fibrosity so that the response is inversely related to the fibrosity of the chip.
In the context of the OrganoPlate$^{\circledR}$, low fibrous or even non-fibrous ECM is desirable, which translates to a high response value.
The main goal of this experiment is to understand which factors are the most influential on the fibrosity, in order to limit the variation in the measured response.

The fibrosity was measured in 256 chips.
However, in 17 of them, the ECM gel did not fill the central channel correctly.
Therefore, these chips were excluded from the analysis.
The final data set thus contains 239 observations, and the response ranges from 244 to 378, with a mean value of 329.
The complete data set containing the fibrosity measurements for all 256 chips, along with the experimental design, is available in the supplementary material of this paper.
The code used for the analysis and to generate the figures is available in the Github repository \url{https://github.com/ABohynDOE/microplate_assays}.

\subsection{Phase I: Identification of the active effects}
Among the eight rows present on all plates, there were two rows that did not contain excluded chips.
We averaged the fibrosity response over these two rows for all 32 columns of the experiment, and use these 32 averages to identify the active factor effects in our analysis.

Table \ref{tab:pse50} shows the effect sizes for fibrosity per stratum, in decreasing order of their absolute value.
For the $Tube$ and $Plate$ strata, the table shows an estimate of the PSE(50) \citep{schoen2000three} robust standard error of the effects, along with a threshold to declare an effect active ($\alpha=10\%$).
Since the $Week$ and $Plate$  strata include only one and two degrees of freedom, respectively, it is not feasible to obtain a robust standard error estimate for these strata and a threshold to identify active effects.
We also see that the robust standard error for the $Tube$ stratum is lower than for the $Unit$ stratum.
This suggests that the random variation between the columns of a plate dominates the variation between the different tubes used within a week.

In view of their size, we tentatively assume that the factors $\mathbf{g}$ and $\mathbf{h}$ and their interaction $\mathbf{gh}$ are active.
The estimates for $\mathbf{cd}$, $\mathbf{a}$, $\mathbf{d}$, $\mathbf{ah}$ and $\mathbf{c}$ are statistically significant in the $Tube$ stratum, and the estimates for $\mathbf{p_5}$ and $\mathbf{p_3}$ are statistically significant in the $Plate$ stratum, implying that the column position is also significant.
Table \ref{tab:aliasing} shows that the significant two-factor interaction $\mathbf{ah}$ is aliased with the two-factor interaction $\mathbf{bc}$.
However, given the effect sizes of the corresponding main effects, $\mathbf{ah}$ is more likely to be active than $\mathbf{bc}$, due to the heredity principle \citep[][Chap. 8]{wu2009experiments}

\input{tables/pse50}

\subsection{Phase II: Model building and data analysis}
Now that we have identified the active effects, we can build a model for the data at the individual chip level.
This has the additional benefit that we can assess the effect of the row positions on the fibrosity.

The experiment was run over two weeks.
We use the index $i$, taking the values 1 or 2, to indicate the weeks.
Within each week, there are two plates and eight tubes.
We use the index $j$, taking the values 1 or 2, to indicate the plates within a week, and the index $k$, taking the values 1 to 8, to indicate the tubes within a week.
However, each plate and tube combination within a week also corresponds to a specific column of the plate (see the full design in Table \ref{tab:full_design_table} as well as Figure \ref{fig:structure_diagram}).
Therefore, the eight columns of a plate are defined by the indices $i$, $j$, and $k$.
Each plate also contains eight rows, for which we use the index $l$, with $l=1,\ldots,8$, to define the eight rows within a plate.
As a result, the 256 chips used in the experiment can be uniquely identified by the indices $i$, $j$, $k$, and $l$, and we can use $Y_{ijkl}$ to denote the fibrosity measured in week $i$, on plate $j$, in column $ijk$ and row $l$, with an ECM gel coming from tube $ik$.
The appropriate model for the fibrosity has the form

\begin{equation}
    Y_{ijkl} = \beta_{0} + \mathbf{x}^{\prime}_{ijkl}\boldsymbol{\beta} + \delta_{i} + \gamma_{ij} + \lambda_{ik} + \phi_{ijk} + \rho_{ijl} + \epsilon_{ijkl}\text{,}
    \label{eq:full-model}
\end{equation}

where $\beta_0$ is a fixed intercept, $\mathbf{x}_{ijkl}$ represents the model expansion of the levels of the treatment factors for chip $ijkl$, $\boldsymbol{\beta}$ is a vector of fixed parameters, $\delta_{i} \sim N\left(0,\sigma_{\delta}^{2}\right)$ is the random effect of the weeks, $\gamma_{ij} \sim N\left(0,\sigma_{\gamma}^{2}\right)$ is the random effect for the plates, $\lambda_{ik} \sim N\left(0,\sigma_{\lambda}^{2}\right)$ is the random effect for the tubes, $\phi_{ijk} \sim N\left(0,\sigma_{\phi}^{2}\right)$ is the random effect for the columns of the plates, $\rho_{ijl} \sim N\left(0,\sigma_{\rho}^{2}\right)$ is a random effect for the rows of the plates and $\epsilon_{ijkl} \sim N\left(0,\sigma^{2}_{\epsilon}\right)$ is the random error of the individual measurement in each chip.
To estimate the variance components of the linear mixed model in Equation \eqref{eq:full-model}, we use restricted maximum likelihood (see \cite{goos2006practical} for more details).

The vector $\mathbf{x}_{ijkl}$, representing the model expansion of the levels of the treatment factors for chip $ijkl$, contains the levels of the factors $\mathbf{a,c,d,g}$ and $\mathbf{h}$ and the cross-products corresponding to the interactions $\mathbf{ah,cd}$ and $\mathbf{gh}$.
To account for the potential edge effects, we also add the main effects of the eight column positions, represented by $\mathbf{p}_1$ to $\mathbf{p}_7$, and the main effects of the eight row positions using $\mathbf{q}_{1}$ to $\mathbf{q}_{7}$.

There are five variance components, two of which ($\sigma^{2}_{\delta}$ and $\sigma^{2}_{\gamma}$) are not estimable because the corresponding strata do not contain enough degrees of freedom.
Estimates of the three estimable variance components, the tube variance $\sigma^{2}_{\lambda}$, the column variance $\sigma_{\phi}^{2}$, and the error variance $\sigma^{2}_{\epsilon}$, are presented in Table \ref{tab:variance}.
The total variance is dominated by the error component, $\sigma^{2}_{\epsilon}$.
This suggests that the major source of random variation is related to measurements at the individual chip level or to the variation among these chips.

\input{tables/variance}

Using F-tests, we compute p-values for the fixed effects; see Table \ref{tab:sign-test}.
Since effects $\mathbf{g}$, $\mathbf{h}$, and $\mathbf{gh}$ are not testable, they are not shown in the table.
We see that, except for the factor $\mathbf{row}$, all effects are significant at the 10 \% level.

\input{tables/sign-test}

\begin{figure}[t]
    \centering
    \includegraphics[width=\textwidth]{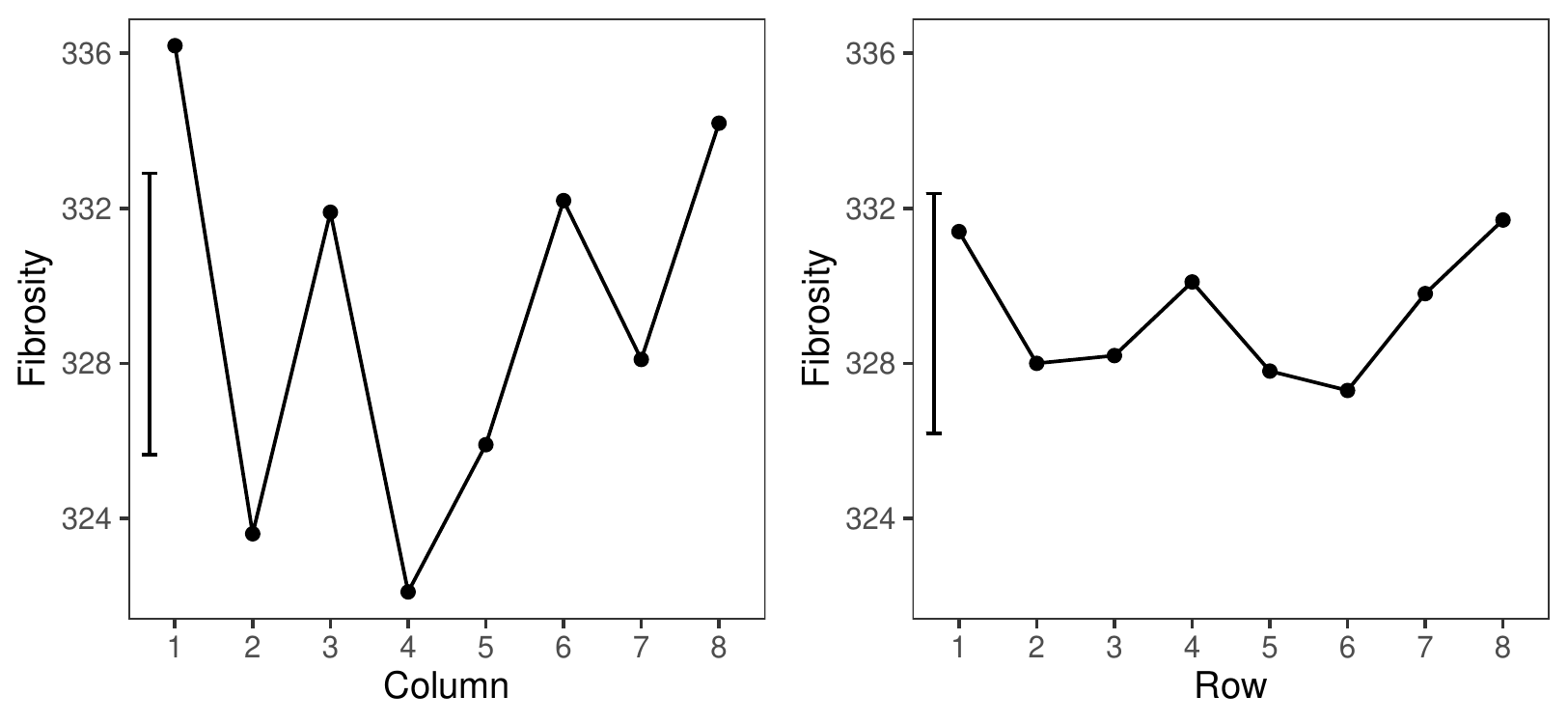}
    \caption{Fitted means of the column and row positions, with an error bar representing the least significant difference $(\alpha=10\%)$.}
    \label{fig:position_plot}
\end{figure}

The small p-value for the column factor and the large p-value for the row factor suggests that an edge effect only seems to occur over the columns of a plate.
To visualize the magnitude of the effects of the row and column positions, we compute the fitted means for each row and column position, along with the least significant difference (LSD) at the 10\% level; see Figure \ref{fig:position_plot}.
None of the pairs of fitted means of the row positions are significantly different.
It confirms that there does not seem to be edge effects over the rows of a plate.
However, there are significant differences between the means of the column positions, although there is no clear division between the edge columns, 2 and 23, and the rest.

\begin{figure*}[hbtp]
	\centering

	\subfloat[$\mathbf{gh}$ interaction]{%
	  \includegraphics[width=0.31\textwidth]{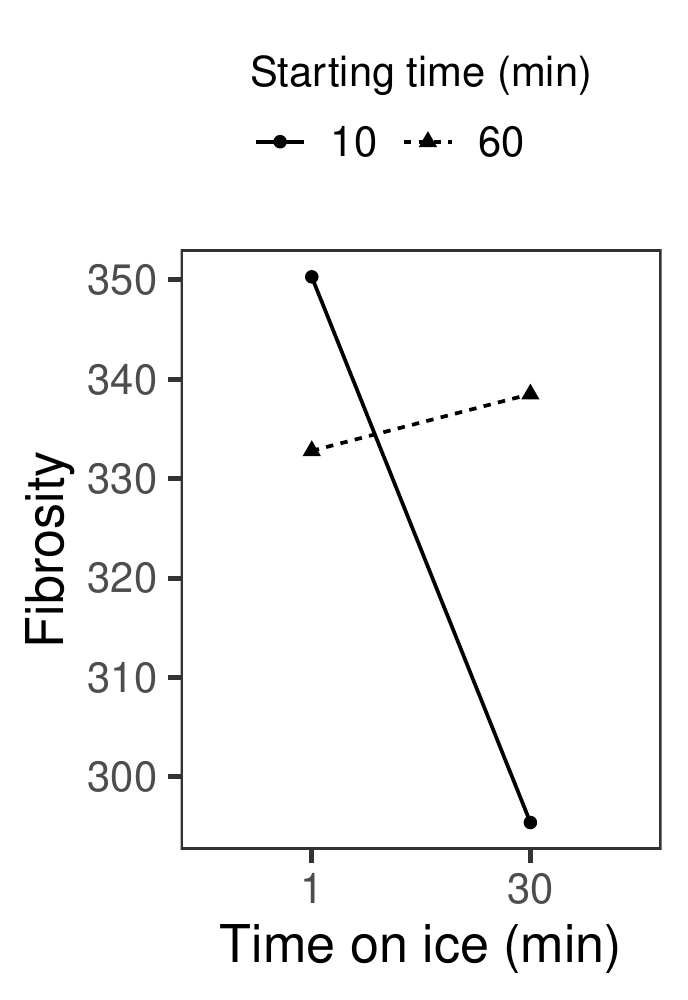}%
	  \label{fig:interaction:gh}%
	}%
	\subfloat[$\mathbf{ah}$ interaction]{%
	  \includegraphics[width=0.31\textwidth]{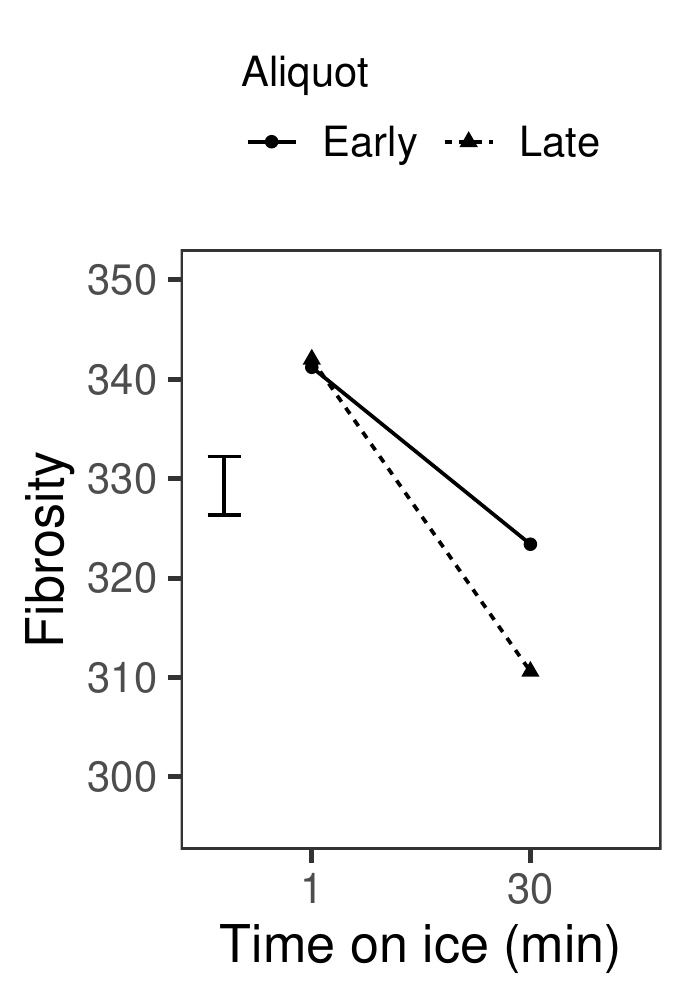}%
	  \label{fig:interaction:ah}%
	}%
	\subfloat[$\mathbf{cd}$ interaction]{%
	  \includegraphics[width=0.31\textwidth]{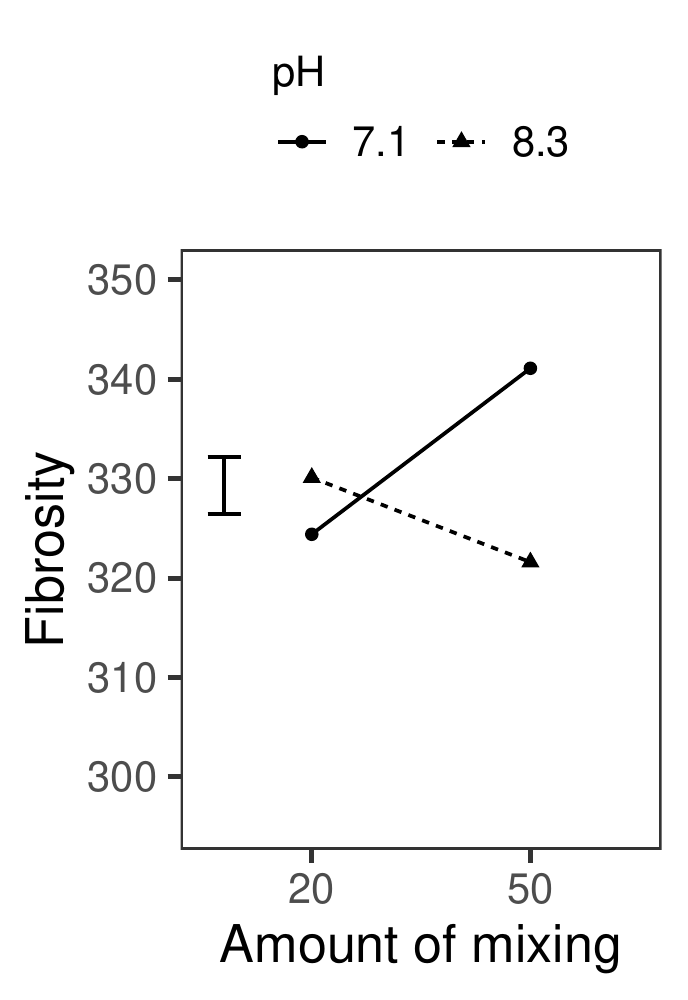}%
	  \label{fig:interaction:cd}%
	}\qquad

	\caption{Interaction plots with their least significant difference $(\alpha=10\%)$ represented by error bars on the left. No LSD can be determined for the first interaction since the variance components $\sigma^{2}_{\delta}$ and $\sigma^{2}_{\gamma}$ cannot be estimated.}
	\label{fig:interaction_plot}
\end{figure*}

Interaction plots for the interactions $\mathbf{cd}$, $\mathbf{ah}$ and $\mathbf{gh}$ are shown in Figure \ref{fig:interaction_plot}, along with the LSD at the 10\% level for $\mathbf{cd}$ and $\mathbf{ah}$.
The first interaction, shown in Figure \ref{fig:interaction:gh}, is $\mathbf{gh}$, i.e., the interaction between starting time and time on ice.
This interaction is clearly the strongest of all three interactions.
However, since it is estimated in the $Plate$ stratum, we cannot determine an LSD for that interaction.
We see that a starting time of ten minutes combined with a time on ice of one minute maximizes the fibrosity response.
However, we see that a starting time of one minute makes the response highly sensitive to the level of time on ice.
So, a starting time of 60 minutes minimizes the effect of time on ice on the fibrosity response.
The second interaction, shown in Figure \ref{fig:interaction:ah}, is $\mathbf{ah}$, i.e., the interaction between aliquot and time on ice.
We see that the level of the factor aliquot only matters if the time on ice is 30 minutes, and that a time on ice of one minute maximizes the fibrosity response, as in Figure \ref{fig:interaction:gh}.
This is in line with the results presented in Table \ref{tab:pse50}, where the estimated effect size of time on ice is strongly negative.

Finally, for the third interaction $\mathbf{cd}$ shown in Figure \ref{fig:interaction:cd}, i.e., the interaction between amount of mixing and pH, we see that a pH of 7.1 in combination with an amount of mixing of 50 stirrings maximizes the fibrosity response but a pH of 8.3 limits the variation in that response.

\subsection{Conclusions on the fibrosity response}
The goal of this study was to detect which factors minimize variation in the fibrosity of the ECM.
A secondary goal specifically for the OrganoPlate$^{\circledR}$ was to establish factor settings with high values of the response (this minimizes fibrosity; see Section \ref{sec:response_variable}).
Based on our analysis, we recommend two treatment combinations, depending on which of the two goals is more important.
To minimize the variation in the fibrosity, use a time on ice of one minute, starting time of 60 minutes, an early aliquot, a pH of 8.3, and mix the preparation 20 times.
To maximize the fibrosity response use a time on ice of one minute, a starting time of ten minutes, an early aliquot, pH of 7.1, and mix the preparation 50 times.
We observed a strong effect of the column positions on the results, which should be taken into account when performing assays.
For this reason, it is important to replicate treatment combinations over different column positions.

\section{Alternative scenarios}\label{sec:alternatives}
\label{sec:alt_scen}
In this section, we consider four alternative scenarios for the experiment.
For each scenario, we justify why it could make sense, describe how the randomization restrictions of the experiment would change, and explain which experimental design we would consider.
We provide the final design for each alternative scenario in the supplementary materials.

The four scenarios considered are:
\begin{enumerate}
	\item Each of the 16 available tube is used twice on a single plate instead of once on each plate within a week.
	Using a tube twice on a single plate creates a split-split-split-plot structure instead of the current split-strip-plot structure.
    \item Thirty-two instead of 16 tubes are available so that each individual column would get its own tube instead of sharing a tube with a column on another plate.
    \item All factors can be varied within the plates. Currently, factors $\mathbf{g}$ and $\mathbf{h}$ are varied between the plates and the weeks, respectively, because of logistic restrictions.
    Without these restrictions, all factors could be varied from column to column.
    \item The blocking factor corresponding to the column positions has four levels.
    It is conceivable that the columns at the same distance of the left- and right-most edges of the plate are alike. In that case, a potential edge effect would be symmetrical and a four-level blocking factor would be sufficient to capture the effect of the column positions.
\end{enumerate}
Before considering the four alternative scenarios, we summarize the process used to generate the experimental design for the actual experiment.

\subsection{Initial design}
The initial design generation process can be summarized in four steps (see Section \ref{sec:trmt_design}):
\begin{enumerate}
    \item Find a $2^{6-1}$ design involving 32 runs for the factors $\mathbf{a}$ to $\mathbf{f}$.
    For this purpose, we chose the minimum aberration design with $\mathbf{f=abcde}$.
    \item Arrange the design in eight orthogonal blocks corresponding to the eight column positions on a plate using the blocking scheme of \cite{xu2006minimum}. In our case, the three defining pseudo-factors were  $\mathbf{p_1=ab}$, $\mathbf{p_2=ce}$ and $\mathbf{p_3=acf}$.
    \item Add a four-level factor orthogonal to factors $\mathbf{a - f}$ and to the eight-level blocking factor, whose two independent pseudo-factors define the two factors $\mathbf{g}$ and $\mathbf{h}$. We considered all options with regular aliasing and selected the one that minimized the aliasing between the interactions among the six two-level factors $\mathbf{a}$ to $\mathbf{f}$, and between the main effects and two-factor interactions of the factors $\mathbf{g}$ and $\mathbf{h}$. In our case, the two additional factors were defined by the relations $\mathbf{g=ace}$ and $\mathbf{h=abc}$.
    \item Ensure that eight tubes per week are used. In our case, this was done by selecting the factor $\mathbf{h}$ to define the weeks.
\end{enumerate}
We now discuss how each of the four steps change when we consider an alternative scenario, along with the allocation of the factorial effects to the different error strata.

\subsection{Scenario 1: tubes are used twice on each plate}
\label{sec:alt_scen_tube_twice}
If each tube were used twice on a single plate, the logistic restrictions concerning the blocks for the eight column positions and the factors $\mathbf{g}$ and $\mathbf{h}$ would still be in place.
This would imply that steps 1 to 3 of the design generation process would be the same as for the initial design, and only the assignment of the tubes to the plates and weeks in Step 4 would change, because we now need to divide the 16 tubes into groups corresponding to the four plates.
We would still use $\mathbf{h=abc}$ to define the weeks, but, instead of using the combinations of factors $\mathbf{a}$, $\mathbf{b}$, $\mathbf{c}$ and $\mathbf{d}$ as the four tube factors (see Table \ref{tab:tube-factor-aliasing}), we would use the factors $\mathbf{a}$, $\mathbf{b}$, $\mathbf{c}$ and $\mathbf{e}$.
Using these four factors to define the tubes would mean that the tubes are confounded with the weeks (defined by $\mathbf{h=abc}$) and with the plates (defined by $\mathbf{g=ace}$), so that the 16 tubes would be divided in four groups of four tubes corresponding to the four plates.

Under this scenario, the $Week$ and $Plate$ strata would still contain the factors $\mathbf{g}$, $\mathbf{h}$ and their interaction $\mathbf{gh}$.
However, the $Tube$ stratum would contain 12 degrees of freedom instead of 14, and the $Unit$ stratum would contain 16 degrees of freedom instead of 14.
This would make the effects in the $Tube$ stratum harder to detect.
The effects contained in the $Tube$ and the $Unit$ strata are presented in the first panel of Table \ref{tab:aliasing_alt_scenarios}, along with the aliasing between the main effects and two-factor interactions of the eight factors $\mathbf{a}$ to $\mathbf{h}$ and the pseudo-factors for the columns positions.

\subsection{Scenario 2: 32 tubes available}
\label{sec:alt_scen_32_tubes}
If 32 tubes were available, the logistic restrictions concerning factors $\mathbf{g}$ and $\mathbf{h}$ would still be in place, and steps 1 to 3 of the design generation process would be the same as for the current design.
Now that 32 tubes are available, we need to ensure that all 16 level combinations of the tube factors $\mathbf{a-d}$ are used within each week.
In the original design, $\mathbf{h=abc}$ was used to define the weeks.
This imposed a restriction on the level combinations of the tube factors $\mathbf{a}$, $\mathbf{b}$ and $\mathbf{c}$, so that half of them were associated with one week and the other half with the other week.
For a design with 32 tubes, we use $\mathbf{g=ace}$ to define the weeks, so that no restriction on level combinations of tube factors is invoked.
The factor $\mathbf{h}$ now defines the plates within a week.

The new arrangement does not change the aliasing of the treatment factors and their interactions, but it changes the randomization structure of the experiment.
As there are now three nested error strata, this is a split-split-plot design.
Indeed, the $Tube$ stratum is merged into the $Unit$ stratum because no factors are varied only at the tube level; they are all varied between weeks, plates, or column positions, respectively.
The table for the aliasing of the main effects, two-factor interactions, and pseudo-factors, and their allocation to the error strata, is similar to Table \ref{tab:aliasing}, but with the columns $Tube$ and $Unit$ merged.
In principle, the 28 effects in the $Unit$ stratum would make for a more efficient robust standard error estimate compared to the case of 14 effects.
However, the effects in that stratum now have a variance that is larger than before, but smaller than the error variance of the effects in the $Tube$ stratum of the original design.
This is an advantage for the detection of effects in the former $Tube$ stratum and a disadvantage for the effects in the former $Unit$ stratum.

\subsection{Scenario 3: no logistic restrictions on the treatment factors}
In this scenario, all factors can be varied from column to column on each plate.
In Step 1, we would consider a $2^{8-3}$ design for factors $\mathbf{a}$ to $\mathbf{h}$.
More specifically, we would use the minimum aberration design with $\mathbf{f=abcd}$, $\mathbf{g=abe}$, and $\mathbf{h=ace}$.
To block the design in Step 2, we would use the blocking scheme of \cite{xu2006minimum} with $\mathbf{p_1 = abc}$, $\mathbf{p_2 = ad}$ and $\mathbf{p_3 = ae}$.
In Step 3, we need to add a four-level factor defining the four plates and the two weeks.
The four levels of the factor can be decomposed in two independent pseudo-factors $\mathbf{b_1}$ and $\mathbf{b_2}$, and we define their interaction as $\mathbf{b_3=b_1b_2}$.
Clearly, we again need to consider all possible ways of adding a four-level factor.
The three options with regular aliasing (effects are either fully aliased or orthogonal) are shown in Table \ref{tab:alt_scenario_aliasing-four-columns-block}.
Option 1 is the best because only three to-factor interactions are aliased with pseudo-factors for the column positions, as opposed to four and six for Options 2 and 3.
Now, the allocation of the 32 treatment combinations to the columns are defined through the eight levels of the eight-level blocking factor, but we still need to assign the experimental runs to the four plates, the two weeks, and the 16 tubes.
To restrict the number of tubes to eight per week, we still need a half-fraction of the $2^{4}$ design for factors $\mathbf{a}$, $\mathbf{b}$, $\mathbf{c}$ and $\mathbf{d}$ inside each week.
Defining the weeks with the second pseudo-factor $\mathbf{b_2=abd}$ ensures that half of the treatment combinations of $\mathbf{a}$, $\mathbf{b}$ and $\mathbf{d}$ occur in one week and the other half in the remaining week.
The eight level combinations of factors $\mathbf{a}$, $\mathbf{b}$, and $\mathbf{d}$ thus define the eight tubes within a week.
Finally, we define the plates using the first pseudo-factor $\mathbf{b_1=ade}$.

This scenario does not change the randomization structure of the experiment.
However, the allocation of the effects to the error strata is different.
The first two strata, namely $Week$ and $Plate$, would contain the interactions $\mathbf{ade}$, $\mathbf{cf}$ and $\mathbf{ag}$ rather than factors $\mathbf{g}$, $\mathbf{h}$ and their interaction $\mathbf{gh}$.
The effects contained in the $Tube$ and $Unit$ strata are presented in the second panel of Table \ref{tab:aliasing_alt_scenarios}, along with the aliasing between the main effects and two-factor interactions of the eight factors $\mathbf{a}$ to $\mathbf{h}$ and the pseudo-factors for the columns positions.
The main advantage of this allocation is that all the main effects of the eight factors are now testable with a robust standard error.

\input{tables/alt_scenario_aliasing_4lvl}

\subsection{Scenario 4: four distinct column positions}
If there were four distinct column positions instead of eight due to the symmetry of the edge effect, the second step of the design process would change because we would need to arrange the design in four blocks instead of eight.
Using the blocking scheme from \cite{xu2006minimum}, the two independent pseudo-factors that create the four-level factor would be $\mathbf{p_1=ab}$ and $\mathbf{p_2=acd}$.
In Step 3, there would again be three regular options to add a four-level factor.
All of these of shown in Table \ref{tab:aliasing-four-columns-block}.
In the first option, interactions $\mathbf{abd}$ and $\mathbf{abc}$ used to define factors $\mathbf{g}$ and $\mathbf{h}$, partition the 16 tubes into four sets of four tubes so that each tube is used twice on a plate.
This option is feasible, but, as Scenario 1 shows, the $Tube$ and $Unit$ strata then have 12 and 16 degrees of freedom, respectively.
We prefer options with equal number of degrees of freedom in each of these strata.

For the second option in Table \ref{tab:aliasing-four-columns-block}, the main effect of factor $\mathbf{h}$ is confounded with the two-factor interaction $\mathbf{ac}$.
This results in a design of resolution III, which is undesirable.
By relabelling the factors, we could use $\mathbf{h=abe}$, $\mathbf{gh=ac}$, and $\mathbf{g=adf}$, which would create a resolution IV design.
However, none of the factors $\mathbf{g}$ or $\mathbf{h}$ could be used to define the weeks where 8 of the 16 tubes are involved.

Finally, the third option in Table \ref{tab:aliasing-four-columns-block} is suitable for our purpose.
It retains a resolution of IV, and, because $\mathbf{h=abc}$, it restricts the number of tubes to 8 per week.
In this scenario, both the randomization structure and the allocation of the factor effects would be similar to the design actually used.
The main difference would be that the column position effects are now modelled with three instead of seven pseudo-factors, so that they are estimated more precisely.

Under this scenario, the $Week$ and $Plate$ strata would still contain the factors $\mathbf{g}$, $\mathbf{h}$ and their interaction $\mathbf{gh}$.
The effects contained in the $Tube$ and the $Unit$ strata are presented in the third panel of Table \ref{tab:aliasing_alt_scenarios}, along with the aliasing between the main effects and two-factor interactions of the eight factors $\mathbf{a}$ to $\mathbf{h}$ and the pseudo-factors for the four columns positions.

\input{tables/aliasing_four_columns_blocking}

\input{tables/aliasing_alternative_scenario}

\subsection{Conclusions for the alternative scenarios}
Each alternative scenario had its own specificity which made it different from the original one.
In Scenario 1, some degrees of freedom were moved from the $Tube$ stratum to the $Unit$ stratum so that, compared with the original scenario, there are more effects estimated in the $Unit$ stratum and less effects in the $Tube$ stratum.
This makes the effects in the Tube stratum harder to detect.
In Scenario 2, the 14 degrees of freedom from both the $Tube$ and the $Unit$ strata were merged so that the tube factors' effects are estimated more precisely than before and the remaining effects in that stratum are estimated less precisely than before.
However the main problem in both these scenarios is that factors $\mathbf{g}$ and $\mathbf{h}$ are fully aliased with the effects of the weeks and the plates.
Scenario 3 would remedy this problem because the main effects of all eight factors are located in strata with 14 degrees of freedom.
They can thus all be evaluated using robust standard errors.
Finally, Scenario 4 provides more information on the factorial effects, because there are only four column positions to account for.
Overall, all four scenarios considered demonstrate how the methodology applied to the initial design changes with the different restrictions.

\section{Conclusions}\label{sec:conclusions}
This paper showed that a small eight-factor experiment can be challenging to design when there are multiple restrictions on the randomization.
While split-plotting, strip-plotting, and blocking are fairly common on their own, the combination of the three in the same design is unusual.
This complicates the generation of the design as there are no catalogs or templates of designs with similar restrictions on the randomization.
We developed a step-by-step approach to create a good experimental design while taking these restrictions into account. This approach allowed us to tackle the problem of including an extra multi-level factor orthogonal to the main randomization structure.

The design allowed us to detect the active effects in the protocol for microplate assays, and to estimate their contribution to the final quality of the microplate.
These insights will be used to improve the protocol for the OrganoPlate$^{\circledR}$.

The alternative scenarios in Section \ref{sec:alt_scen} should be helpful for practitioners to apply the step by step approach for building in multiple restrictions in the randomization in their own experiments.

\bigskip
\begin{center}
{\large\bf Supplementary material}
\end{center}

\begin{description}
\item[data\_table.csv:] File containing the fibrosity measurement data for all 256 chips.
\item[alternative\_scenarios.xlsx:] Spreadsheet containing the final design chosen for each of the four alternative scenarios described in Section \ref{sec:alt_scen}. (Excel file)

\end{description}

\if0\blind
{
\bigskip
\begin{center}
{\large\bf Acknowledgements}
\end{center}
This research was financially supported by the Fonds Wetenschappelijk Onderzoek  (FWO, Flanders, Belgium).
} \fi

\bibliographystyle{apalike} 
\bibliography{references} 


\end{document}

%% file: tables/factor_levels.tex
\begin{table}[htbp]
  \centering
  \caption{Experimental factors and their settings.}
    \begin{tabular}{lcllll}
        \toprule
        Stage                                                   & Label             & Factor                    & \multicolumn{2}{c}{Settings}\\
        \midrule
        \multicolumn{1}{l}{\multirow{4}[1]{*}{Mixing}}          & $\mathbf{a}$      & Aliquot                                  & Early & Late \\
                                                                & $\mathbf{b}$      & Collagen gel volume (µL)                 & 100   & 300  \\
                                                                & $\mathbf{c}$      & Amount of mixing                         & 20    & 50   \\
                                                                & $\mathbf{d}$      & pH of solution                           & 7.1   & 8.3  \\
        \midrule
        \multicolumn{1}{l}{\multirow{2}[1]{*}{Timing}}          & $\mathbf{g}$      & Starting time (min)                      & 10    & 60   \\
                                                                & $\mathbf{h}$      & Time on ice (min)                        & 1     & 30   \\
        \midrule
        \multicolumn{1}{l}{\multirow{2}[1]{*}{Post-Loading}}    & $\mathbf{e}$      & HBSS \ce{Ca++} and \ce{Mg++} content     & Low   & High  \\
                                                                & $\mathbf{f}$      & HBSS removal                             & No    & Yes   \\
        \bottomrule
    \end{tabular}
  \label{tab:factor_levels}
\end{table}

%% file: figures/structure_diagram.tex
\begin{tabular}{|c|c|c|c|c|c|c|c|c|c|c|c|c|c|c|c|c|c|}
    \hline
    \multirow{2}[4]{*}{Week}    & \multirow{2}[4]{*}{Plate} & \multicolumn{16}{c|}{Tube} \\
    \cline{3-18}                &                           & 1 & 2 & 3 & 4 & 5 & 6 & 7 & 8 & 9 & 10 & 11 & 12 & 13 & 14 & 15 & 16 \\
    \hline
    \multirow{2}[4]{*}{1}       & 1                         & $\bullet$ & $\bullet$ & $\bullet$ & $\bullet$ & $\bullet$ & $\bullet$ & $\bullet$ & $\bullet$ &   &   &   &   &   &   &   &  \\
    \cline{2-18}                & 2                         & $\bullet$ & $\bullet$ & $\bullet$ & $\bullet$ & $\bullet$ & $\bullet$ & $\bullet$ & $\bullet$ &   &   &   &   &   &   &   &  \\
    \hline
    \multirow{2}[4]{*}{2}       & 3                         &   &   &   &   &   &   &   &   & $\bullet$ & $\bullet$ & $\bullet$ & $\bullet$ & $\bullet$ & $\bullet$ & $\bullet$ & $\bullet$ \\
    \cline{2-18}                & 4                         &   &   &   &   &   &   &   &   & $\bullet$ & $\bullet$ & $\bullet$ & $\bullet$ & $\bullet$ & $\bullet$ & $\bullet$ & $\bullet$ \\
    \hline
\end{tabular}

%% file: tables/aliasing-eight-level-factor.tex
\begin{table}[htbp]
    \centering
    \caption{Aliasing of the pseudo-factors $\mathbf{p_i}$  ($i=1,\ldots,7$), defining the column positions, with interactions among the six two-level factors $\mathbf{a} - \mathbf{f}$ in the $2^{6-1}$ design.}
    \begin{tabular}{ll}
        \toprule
        Pseudo-factor   & \multicolumn{1}{c}{Aliasing} \\
        \midrule
        $\mathbf{p_1}$                & $\mathbf{ab}$\\
        $\mathbf{p_2}$                & $\mathbf{ce}$\\
        $\mathbf{p_3}$                & $\mathbf{acf + bde}$ \\
        $\mathbf{p_4}$                & $\mathbf{df}$ \\
        $\mathbf{p_5}$                & $\mathbf{ade + bcf}$ \\
        $\mathbf{p_6}$                & $\mathbf{aef + bcd}$ \\
        $\mathbf{p_7}$                & $\mathbf{acd + bef}$ \\
        \bottomrule
    \end{tabular}%
    \label{tab:aliasing-eight-level-factor}%
\end{table}%

%% file: tables/grouping-scheme.tex
\begin{table}[htbp]
	\centering
	\caption{Grouping scheme used to assign treatment combinations to the eight column positions of a plate.}
	\begin{tabular}{cccc}
		\toprule
		$\mathbf{p_1}=\mathbf{ab}$ 	& $\mathbf{p_2}=\mathbf{ce}$ 	& $\mathbf{p_3}=\mathbf{acf}$ 	& Column position 	\\
		\midrule
		$-$                         & $+$                           & $-$                           & 1 \\
	    $-$                         & $-$                           & $+$                           & 2 \\
	    $-$                         & $-$                           & $-$                           & 3 \\
	    $-$                         & $+$                           & $+$                           & 4 \\
	    $+$                         & $-$                           & $-$                           & 5 \\
	    $+$                         & $-$                           & $+$                           & 6 \\
	    $+$                         & $+$                           & $-$                           & 7 \\
	    $+$                         & $+$                           & $+$                           & 8 \\
		\bottomrule
	\end{tabular}%
	\label{tab:grouping-scheme-eight-level}%
\end{table}%

%% file: tables/aliasing-four-level-factor.tex
\begin{table}[!t]
    \centering
    \caption{Aliasing of factors $\mathbf{g}$ and $\mathbf{h}$ and their interaction $\mathbf{gh}$ for each of the three options for adding a four-level factor to the $2^{6-1}$ design.}
    \begin{tabular}{ccr}
        \toprule
        Option                      & \multicolumn{1}{c}{Factor}    & \multicolumn{1}{c}{Aliasing} \\
        \midrule
        \multirow{3}[0]{*}{1}       &  $\mathbf{g}$                 & $\mathbf{ace+bdf}$ \\
                                    &  $\mathbf{h}$                 & $\mathbf{abc+def}$ \\
                                    &  $\mathbf{gh}$                & $\mathbf{be}$ \\
        \midrule
        \multirow{3}[0]{*}{2}       &  $\mathbf{g}$                 & $\mathbf{cd}$ \\
                                    &  $\mathbf{h}$                 & $\mathbf{ad}$ \\
                                    &  $\mathbf{gh}$                & $\mathbf{ac}$ \\
        \midrule
        \multirow{3}[0]{*}{3}       &  $\mathbf{g}$                 & $\mathbf{ef}$ \\
                                    &  $\mathbf{h}$                 & $\mathbf{ad}$ \\
                                    &  $\mathbf{gh}$                & $\mathbf{ce}$ \\
        \bottomrule
    \end{tabular}%
    \label{tab:aliasing-four-level-factor}%
\end{table}%

%% file: tables/tube-factor-aliasing.tex
\begin{table}[tbp]
    \centering
    \caption{Definition of the tube numbers using the coded levels of the factors $\mathbf{a}$, $\mathbf{b}$, $\mathbf{c}$ and $\mathbf{d}$. The vertical dashed line shows the two subdesigns used in the two weeks.}
    \begin{tabular}{lcccccccc|cccccccc}
        \hline
        Week & \multicolumn{8}{c|}{1}        & \multicolumn{8}{c}{2} \\
        \hline
        Tube & 1 & 2 & 3 & 4 & 5 & 6 & 7 & 8 & 9 & 10 & 11 & 12 & 13 & 14 & 15 & 16 \\
        \hline
        $\mathbf{a}$ & $-$ & $-$ & $-$ & $-$ & $+$ & $+$ & $+$ & $+$ & $-$ & $-$ & $-$ & $-$ & $+$ & $+$ & $+$ & $+$ \\
        $\mathbf{b}$ & $-$ & $-$ & $+$ & $+$ & $-$ & $-$ & $+$ & $+$ & $-$ & $-$ & $+$ & $+$ & $-$ & $-$ & $+$ & $+$ \\
        $\mathbf{c}$ & $-$ & $-$ & $+$ & $+$ & $+$ & $+$ & $-$ & $-$ & $+$ & $+$ & $-$ & $-$ & $-$ & $-$ & $+$ & $+$ \\
        $\mathbf{d}$ & $-$ & $+$ & $-$ & $+$ & $-$ & $+$ & $-$ & $+$ & $-$ & $+$ & $-$ & $+$ & $-$ & $+$ & $-$ & $+$ \\
        \hline
        $\mathbf{h}$ & \multicolumn{8}{c|}{$-$}      & \multicolumn{8}{c}{$+$} \\
        \hline
    \end{tabular}%
  \label{tab:tube-factor-aliasing}%
\end{table}%

%% file: tables/full_design_table.tex
\begin{table}[hbtp]
\centering
\caption{Final 32-run experimental design, along with the corresponding week, plate, tube, and column for each experimental run.} 
\label{tab:full_design_table}
\begin{tabular}{cccccccccccc}
  \toprule
Week & Plate & Column & Tube & $\mathbf{ a }$ & $\mathbf{ b }$ & $\mathbf{ c }$ & $\mathbf{ d }$ & $\mathbf{ e }$ & $\mathbf{ f }$ & $\mathbf{ g }$ & $\mathbf{ h }$ \\ 
  \midrule
  1 & 1 & 1 & 3 & $-$ & $+$ & $+$ & $-$ & $+$ & $+$ & $-$ & $-$ \\ 
  1 & 1 & 2 & 6 & $+$ & $-$ & $+$ & $+$ & $-$ & $+$ & $-$ & $-$ \\ 
  1 & 1 & 3 & 5 & $+$ & $-$ & $+$ & $-$ & $-$ & $-$ & $-$ & $-$ \\ 
  1 & 1 & 4 & 4 & $-$ & $+$ & $+$ & $+$ & $+$ & $-$ & $-$ & $-$ \\ 
  1 & 1 & 5 & 7 & $+$ & $+$ & $-$ & $-$ & $+$ & $+$ & $-$ & $-$ \\ 
  1 & 1 & 6 & 8 & $+$ & $+$ & $-$ & $+$ & $+$ & $-$ & $-$ & $-$ \\ 
  1 & 1 & 7 & 1 & $-$ & $-$ & $-$ & $-$ & $-$ & $-$ & $-$ & $-$ \\ 
  1 & 1 & 8 & 2 & $-$ & $-$ & $-$ & $+$ & $-$ & $+$ & $-$ & $-$ \\
  \midrule
  1 & 2 & 1 & 6 & $+$ & $-$ & $+$ & $+$ & $+$ & $-$ & $+$ & $-$ \\ 
  1 & 2 & 2 & 3 & $-$ & $+$ & $+$ & $-$ & $-$ & $-$ & $+$ & $-$ \\ 
  1 & 2 & 3 & 4 & $-$ & $+$ & $+$ & $+$ & $-$ & $+$ & $+$ & $-$ \\ 
  1 & 2 & 4 & 5 & $+$ & $-$ & $+$ & $-$ & $+$ & $+$ & $+$ & $-$ \\ 
  1 & 2 & 5 & 2 & $-$ & $-$ & $-$ & $+$ & $+$ & $-$ & $+$ & $-$ \\ 
  1 & 2 & 6 & 1 & $-$ & $-$ & $-$ & $-$ & $+$ & $+$ & $+$ & $-$ \\ 
  1 & 2 & 7 & 8 & $+$ & $+$ & $-$ & $+$ & $-$ & $+$ & $+$ & $-$ \\ 
  1 & 2 & 8 & 7 & $+$ & $+$ & $-$ & $-$ & $-$ & $-$ & $+$ & $-$ \\ 
  \midrule
  2 & 3 & 1 & 12 & $-$ & $+$ & $-$ & $+$ & $-$ & $-$ & $-$ & $+$ \\ 
  2 & 3 & 2 & 13 & $+$ & $-$ & $-$ & $-$ & $+$ & $-$ & $-$ & $+$ \\ 
  2 & 3 & 3 & 14 & $+$ & $-$ & $-$ & $+$ & $+$ & $+$ & $-$ & $+$ \\ 
  2 & 3 & 4 & 11 & $-$ & $+$ & $-$ & $-$ & $-$ & $+$ & $-$ & $+$ \\ 
  2 & 3 & 5 & 16 & $+$ & $+$ & $+$ & $+$ & $-$ & $-$ & $-$ & $+$ \\ 
  2 & 3 & 6 & 15 & $+$ & $+$ & $+$ & $-$ & $-$ & $+$ & $-$ & $+$ \\ 
  2 & 3 & 7 & 10 & $-$ & $-$ & $+$ & $+$ & $+$ & $+$ & $-$ & $+$ \\ 
  2 & 3 & 8 & 9 & $-$ & $-$ & $+$ & $-$ & $+$ & $-$ & $-$ & $+$ \\ 
  \midrule
  2 & 4 & 1 & 13 & $+$ & $-$ & $-$ & $-$ & $-$ & $+$ & $+$ & $+$ \\ 
  2 & 4 & 2 & 12 & $-$ & $+$ & $-$ & $+$ & $+$ & $+$ & $+$ & $+$ \\ 
  2 & 4 & 3 & 11 & $-$ & $+$ & $-$ & $-$ & $+$ & $-$ & $+$ & $+$ \\ 
  2 & 4 & 4 & 14 & $+$ & $-$ & $-$ & $+$ & $-$ & $-$ & $+$ & $+$ \\ 
  2 & 4 & 5 & 9 & $-$ & $-$ & $+$ & $-$ & $-$ & $+$ & $+$ & $+$ \\ 
  2 & 4 & 6 & 10 & $-$ & $-$ & $+$ & $+$ & $-$ & $-$ & $+$ & $+$ \\ 
  2 & 4 & 7 & 15 & $+$ & $+$ & $+$ & $-$ & $+$ & $-$ & $+$ & $+$ \\ 
  2 & 4 & 8 & 16 & $+$ & $+$ & $+$ & $+$ & $+$ & $+$ & $+$ & $+$ \\ 
   \bottomrule
\end{tabular}
\end{table}

%% file: tables/total_aliasing.tex
\begin{table}[htbp]
  \centering
  \caption{Aliasing of main effects, two-factor interactions, some three-factor interactions and pseudo-factors of the column positions, as well as the stratum in which each of the effects is estimated, and the total number of degrees of freedom per stratum.}
    \begin{tabular}{llll}
    \toprule
    $Week$  & $Plate$ & $Tube$ & $Unit$ \\
    \midrule
    $\mathbf{h}  $   	& $\mathbf{g}  $     	& $\mathbf{a}  $          		& $\mathbf{e}  $ \\
                        & $\mathbf{gh + be}$    & $\mathbf{b}  $          		& $\mathbf{f}  $ \\
                        &                       & $\mathbf{c}  $          		& $\mathbf{p_2 + ag + ce}$  \\
                        &                       & $\mathbf{d}  $          		& $\mathbf{p_3}  $ \\
                        &                       & $\mathbf{p_1 + ab + ch}$   	& $\mathbf{p_4 + bg + df + eh }$  \\
                        &                       & $\mathbf{p_6}  $        		& $\mathbf{p_5}  $ \\
                        &                       & $\mathbf{p_7}  $        		& $\mathbf{ae + cg}$  \\
                        &                       & $\mathbf{ac + bh + eg}$    	& $\mathbf{af}$ \\
                        &                       & $\mathbf{ad}$              	& $\mathbf{bf + dg}$  \\
                        &                       & $\mathbf{ah + bc}$         	& $\mathbf{cf}$ \\
                        &                       & $\mathbf{bd + fg}$         	& $\mathbf{de + fh}$  \\
                        &                       & $\mathbf{cd}$              	& $\mathbf{agh}$ \\
                        &                       & $\mathbf{dh + ef}$         	& $\mathbf{adg}$ \\
                        &						& $\mathbf{abd}$				& $\mathbf{cgh}$ \\
    \midrule
    1 df			& 2 df 				& 14 df						& 14 df  \\
    \bottomrule
    \end{tabular}%
  \label{tab:aliasing}%
\end{table}%

%% file: tables/pse50.tex
\begin{table}[hbtp!]
    \caption{Effect sizes for fibrosity per stratum. The effects are arranged in decreasing order of absolute size. For the $Tube$ and $Unit$ strata, the figures in parentheses are the robust standard error and the critical value corresponding to a $10\%$ significance level.}
    \begin{minipage}[t]{.33\textwidth}
        \caption*{}
        \begin{tabular}{lr}
        \toprule
        \multicolumn{2}{l}{$Week$}       \\
        \midrule
        Effect & Estimate  \\
        \midrule
        $\mathbf{h}$                        & $-13.83$    \\
        \midrule
        \multicolumn{2}{l}{$Plate$} \\
        \midrule
        Effect & Estimate  \\
        \midrule
        $\mathbf{gh}$	                    & $16.27$      \\   
    	$\mathbf{g}$                        & $5.39$ 	     \\
    	\bottomrule
		\end{tabular}%
    \end{minipage}%
    \hfill
    \begin{minipage}[t]{.33\textwidth}
        \caption*{}
        \begin{tabular}{lrrp{1cm}}
        \toprule
        \multicolumn{2}{l}{$Tube$ $(0.9, 1.55)$} \\
        \midrule
        Effect & Estimate \\
        \midrule
        $\mathbf{cd}$	                    & $-5.45$  \\
    	$\mathbf{a}$                        & $-4.89$  \\
    	$\mathbf{d}$                        & $-4.55$  \\
    	$\mathbf{ah}$	                    & $-3.02$  \\       
    	$\mathbf{c}$                        & $2.27	$  \\
    	$\mathbf{fg}$	                    & $-1.45$ \\
    	$\mathbf{p_7}$ 	                    & $1.42	$ \\
    	$\mathbf{b}$                        & $0.77	$ \\
	    $\mathbf{eg}$	                    & $-0.61$ \\
    	$\mathbf{ad}$	                    & $0.58	$ \\
    	$\mathbf{p_1}$	                    & $-0.55$ \\
    	$\mathbf{dh}$	                    & $0.39	$ \\
    	$\mathbf{p_6}$ 	                    & $-0.30$ \\
    	$\mathbf{abd}$ 	                    & $-0.27$ \\
    	\bottomrule
    	\end{tabular}
    \end{minipage}
    \hfill
    \begin{minipage}[t]{.33\textwidth}
        \caption*{}
        \begin{tabular}{lrrp{1cm}}
        \toprule
		\multicolumn{2}{l}{$Unit$ $(1.32, 2.26)$}  \\
		\midrule
        Effect & Estimate \\
        \midrule
        $\mathbf{p_5}$ 	& $5.20$ \\
		$\mathbf{p_3}$ 	& $-2.77$ \\
		$\mathbf{cgh}$ 	& $-1.95$ \\
		$\mathbf{cg}$	& $-1.33$ \\
		$\mathbf{cf}$	& $-1.27$ \\
		$\mathbf{agh}$ 	& $1.02 $ \\
		$\mathbf{adg}$ 	& $0.92 $ \\
		$\mathbf{fh}$	& $0.89 $ \\
		$\mathbf{f}$    & $-0.80$ \\
		$\mathbf{p_2}$	& $0.70 $ \\
		$\mathbf{p_4}$	& $-0.58$ \\
		$\mathbf{af}$	& $-0.42$ \\
		$\mathbf{dg}$	& $-0.39$ \\
		$\mathbf{e}$    & $-0.14$ \\
        \bottomrule		
        \end{tabular}%
		\vfill
    \end{minipage}
    \label{tab:pse50}
\end{table}

%% file: tables/variance.tex
\begin{table}[hbtp]
    \centering
    \caption{Point estimates and standard errors of the five variance components in the model in Equation \eqref{eq:full-model}.}
    \begin{tabular}{lcrr}
        \toprule
        Stratum     & Variance component        & Estimate & Standard error \\
        \midrule
        Week        & $\sigma^2_{\delta}$       & $-$   & $-$            \\
        Plate       & $\sigma^2_{\gamma}$       & $-$   & $-$            \\
        Tube        & $\sigma^2_{\lambda}$      & 2.5   & 12.1        \\
        Column      & $\sigma^2_{\phi}$         & 6.2   & 13.6        \\
        Row         & $\sigma^2_{\rho}$         & 1.9   & 8.3       \\
        Error       & $\sigma^2_{\epsilon}$     & 179.8 & 18.7         \\
        \bottomrule
    \end{tabular}
    \label{tab:variance}
\end{table}

%% file: tables/sign-test.tex
\begin{table}[htbp]
  \centering
  \caption{Significance test for the fixed effects.}
    \begin{tabular}{lcccc}
    \toprule
    Effect              & DF numerator   & DF denominator   & F-statistic   & p-value \\
    \midrule    
    $\mathbf{a}$        & 1         & 5.8       & 8.59          & 0.027 \\
    $\mathbf{c}$        & 1         & 5.9       & 3.88          & 0.097 \\
    $\mathbf{d}$        & 1         & 5.9       & 10.97         & 0.017 \\
    $\mathbf{cd}$       & 1         & 5.9       & 36.13         & 0.001 \\
    $\mathbf{ah}$       & 1         & 5.8       & 10.71         & 0.018 \\
    $\mathbf{row}$      & 7         & 20.1      & 0.46          & 0.850 \\
    $\mathbf{column}$   & 7         & 9.9       & 3.17          & 0.049 \\
    \bottomrule
    \end{tabular}%
  \label{tab:sign-test}%
\end{table}%

%% file: tables/alt_scenario_aliasing_4lvl.tex
\begin{table}[htbp]
    \centering
    \caption{Aliasing of the three pseudo-factors ($\mathbf{b_i}$ with $i=1,\ldots,3$) defining the weeks and the plates with interactions among the eight factors of the $2^{8-3}$ design for three blocking options in Scenario 3.}
    \begin{tabular}{cll}
        \toprule
        Option                  & Factor            & Aliasing \\
        \midrule
        \multirow{3}[0]{*}{1}   &  $\mathbf{b_1}$           & $\mathbf{ade + bdg + cdh}$ \\
                                &  $\mathbf{b_2}$           & $\mathbf{cf + abd + deg}$        \\
                                &  $\mathbf{b_3}$           & $\mathbf{ag + be + dfh}$ \\
        \midrule
        \multirow{3}[0]{*}{2}   &  $\mathbf{b_1}$           & $\mathbf{bc + gh + adf}$\\
                                &  $\mathbf{b_2}$           & $\mathbf{cf + abd + deg}$ \\
                                &  $\mathbf{b_3}$           & $\mathbf{bf + acd + deh}$ \\
        \midrule
        \multirow{3}[0]{*}{3}   &  $\mathbf{b_1}$           & $\mathbf{bc + gh + adf}$ \\
                                &  $\mathbf{b_2}$           & $\mathbf{ac + eh + bdf}$ \\
                                &  $\mathbf{b_3}$           & $\mathbf{ab + eg + cdf}$ \\
        \bottomrule
    \end{tabular}%
    \label{tab:alt_scenario_aliasing-four-columns-block}%
\end{table}%

%% file: tables/aliasing_four_columns_blocking.tex
\begin{table}[t]
    \centering
    \caption{Aliasing of the two two-level factors $\mathbf{g}$ and $\mathbf{h}$, and their interaction 
    $\mathbf{gh}$, for each of the three options for adding a four-level factor to the $2^{6-1}$ design in Scenario 4, where a four-level factor suffices to model column positions.}
    \begin{tabular}{crr}
        \toprule
        Option                  & Factor            & Aliasing \\
        \midrule
        \multirow{3}[0]{*}{1}   &  $\mathbf{g}$     & $\mathbf{abd + cef}$ \\
                                &  $\mathbf{h}$     & $\mathbf{abc + def}$ \\
                                &  $\mathbf{gh}$    & $\mathbf{cd}$        \\
        \midrule
        \multirow{3}[0]{*}{2}   &  $\mathbf{g}$     & $\mathbf{adf + bce}$\\
                                &  $\mathbf{h}$     & $\mathbf{ac}$ \\
                                &  $\mathbf{gh}$    & $\mathbf{abe + cdf}$ \\
        \midrule
        \multirow{3}[0]{*}{3}   &  $\mathbf{g}$     & $\mathbf{abe + cdf}$ \\
                                &  $\mathbf{h}$     & $\mathbf{abc + def}$ \\
                                &  $\mathbf{gh}$    & $\mathbf{ce}$ \\
        \bottomrule
    \end{tabular}%
    \label{tab:aliasing-four-columns-block}%
\end{table}%

%% file: tables/aliasing_alternative_scenario.tex
\begin{landscape}
\begin{table}[t]
    \caption{Aliasing of main effects, two-factor interactions,  some three-factor interactions and pseudo-factors of the column positions, in the $Tube$ and $Unit$ strata, for Scenarios 1, 3, and 4.}
    \begin{minipage}[b]{.33\textwidth}
        \centering
        \caption*{Scenario 1: each tube is used twice on a single plate.}
        \begin{tabular}{ll}
        \toprule
        $Tube$                          & $Unit$ \\
        \midrule
        $\mathbf{a  }$                  & $\mathbf{d}$ \\
        $\mathbf{b  }$                  & $\mathbf{f}$ \\
        $\mathbf{c  }$                  & $\mathbf{p_3}$ \\
        $\mathbf{e  }$                  & $\mathbf{p_5}$ \\
        $\mathbf{p_1 + ab + ch}$        & $\mathbf{p_6}$ \\
        $\mathbf{p_2 + ag + ce}$        & $\mathbf{p_7}$ \\
        $\mathbf{p_4 + bg + df + eh}$   & $\mathbf{ad}$ \\
        $\mathbf{ac + bh + eg }$        & $\mathbf{af}$ \\
        $\mathbf{ae + cg}$              & $\mathbf{bd + fg}$ \\
        $\mathbf{ah + bc}$              & $\mathbf{bf + dg}$ \\
        $\mathbf{abe}$                  & $\mathbf{cd}$ \\
        $\mathbf{abg}$                  & $\mathbf{cf}$ \\
                                        & $\mathbf{de + fh}$ \\
                                        & $\mathbf{dh + ef}$ \\
                                        & $\mathbf{abd}$ \\
                                        & $\mathbf{abf}$ \\
   		\midrule
   		12 df 							& 16 df \\
        \bottomrule
    \end{tabular}%
    \end{minipage}%
    \hfill%
    \begin{minipage}[b]{.33\textwidth}
        \centering
        \caption*{Scenario 3: all factors can be varied within plates.}
        \begin{tabular}{ll}
		\toprule
	    $Tube$ & $Unit$ \\
		\midrule
		$\mathbf{a}$        & $\mathbf{e}$ \\
		$\mathbf{b}$        & $\mathbf{g}$	\\
		$\mathbf{c}$        & $\mathbf{h}$ \\
		$\mathbf{d}$        & $\mathbf{p_3 + ae + bg + ch}$  \\
		$\mathbf{f}$        & $\mathbf{p_5}$ \\
		$\mathbf{p_1 + df}$  & $\mathbf{p_6 + de}$ \\
		$\mathbf{p_2 + ad}$  & $\mathbf{p_7 + ef}$ \\
		$\mathbf{p_4 + af}$  & $\mathbf{ah + ce}$ \\
		$\mathbf{ab + eg}$  & $\mathbf{bh + cg}$ \\
		$\mathbf{ac + eh}$  & $\mathbf{dg }$\\
		$\mathbf{bc + gh}$  & $\mathbf{dh }$\\
		$\mathbf{bd}$       & $\mathbf{fg }$\\
		$\mathbf{bf}$       & $\mathbf{fh }$\\
		$\mathbf{cd}$       & $\mathbf{aef}$\\
		\midrule
   		14 df 				& 14 df \\
        \bottomrule
		\end{tabular}%
    \end{minipage}%
    \hfill
    \begin{minipage}[b]{.43\textwidth}
        \centering
        \caption*{Scenario 4: the blocking factor corresponding to the column positions has four levels.}
        \begin{tabular}{ll}
		\toprule
		$Tube$ & $Unit$ \\
		\midrule
		 $\mathbf{a}$ 				& $\mathbf{e}$ \\
		 $\mathbf{b}$ 				& $\mathbf{f}$ \\
		 $\mathbf{c}$ 				& $\mathbf{ae + bg}$ \\
		 $\mathbf{d}$ 				& $\mathbf{af}$ \\
		 $\mathbf{p1 + ab + ch + eg}$ & $\mathbf{ag + be}$ \\
		 $\mathbf{p2}$ 				& $\mathbf{bf}$ \\
		 $\mathbf{p3}$ 				& $\mathbf{cf + dg}$ \\
		 $\mathbf{ac + bh}$ 			& $\mathbf{cg + df + eh}$ \\
		 $\mathbf{ad}$ 				& $\mathbf{de + fh}$ \\
		 $\mathbf{ah + bc}$ 			& $\mathbf{abf }$ \\
		 $\mathbf{bd}$ 				& $\mathbf{ace }$ \\
		 $\mathbf{cd}$ 				& $\mathbf{acf }$ \\
		 $\mathbf{dh + ef}$ 			& $\mathbf{acg }$\\
		 $\mathbf{abd }$       		& $\mathbf{ade }$ \\
		\midrule
   		14 df 				& 14 df \\
        \bottomrule
		\end{tabular}%
    \end{minipage}
    \hfill
    \label{tab:aliasing_alt_scenarios}
\end{table}
\end{landscape}